\documentclass{styles/svproc}

\usepackage{url}

\usepackage{graphicx}
\usepackage{booktabs}
\usepackage{amsmath,amssymb,amsfonts}
\usepackage{array, multirow, boldline}
\usepackage{subcaption}
\usepackage{hyperref}

\usepackage{orcidlink}

\begin{document}
\mainmatter              
\title{MetaFormer-driven Encoding Network for Robust Medical Semantic Segmentation}
\titlerunning{MFEnNet for Robust Medical Semantic Segmentation}  

\author{
Le-Anh Tran
\and Chung Nguyen Tran \and \\
Nhan Cach Dang 
\and Anh Le Van Quoc 
\and Jordi Carrabina \and \\
David Castells-Rufas 
\and Minh Son Nguyen}

\authorrunning{Le-Anh Tran et al.} 

\institute{}

\maketitle              

\begin{abstract}
Semantic segmentation is crucial for medical image analysis, enabling precise disease diagnosis and treatment planning. However, many advanced models employ complex architectures, limiting their use in resource-constrained clinical settings. This paper proposes MFEnNet, an efficient medical image segmentation framework that incorporates MetaFormer in the encoding phase of the U-Net backbone. MetaFormer, an architectural abstraction of vision transformers, provides a versatile alternative to convolutional neural networks by transforming tokenized image patches into sequences for global context modeling. To mitigate the substantial computational cost associated with self-attention, the proposed framework replaces conventional transformer modules with pooling transformer blocks, thereby achieving effective global feature aggregation at reduced complexity. In addition, Swish activation is used to achieve smoother gradients and faster convergence, while spatial pyramid pooling is incorporated at the bottleneck to improve multi-scale feature extraction. Comprehensive experiments on different medical segmentation benchmarks demonstrate that the proposed MFEnNet approach attains competitive accuracy while significantly lowering computational cost compared to state-of-the-art models. The source code for this work is available at \href{https://github.com/tranleanh/mfennet}{https://github.com/tranleanh/mfennet}.

\keywords{medical image segmentation, semantic segmentation, U-Net, vision transformer, metaformer}
\end{abstract}

\section{Introduction}

Semantic segmentation of medical images is a cornerstone task in computer-assisted diagnosis, treatment planning, and surgical navigation. Precise delineation of anatomical and pathological regions supports reliable clinical decisions and tasks like disease classification. Given its centrality, extensive research has been devoted to developing segmentation models that strike a balance between precision and computational efficiency. Early convolutional neural networks (CNNs), such as U-Net \cite{ronneberger2015u} and its variants, have been widely adopted for medical imaging due to their ability to capture local patterns and adapt to various imaging modalities. However, CNNs struggle with long-range dependencies, limiting their capability to segment complex or diffuse structures accurately.

\begin{figure*}[t]
\centering
\includegraphics[width=1.0\textwidth]{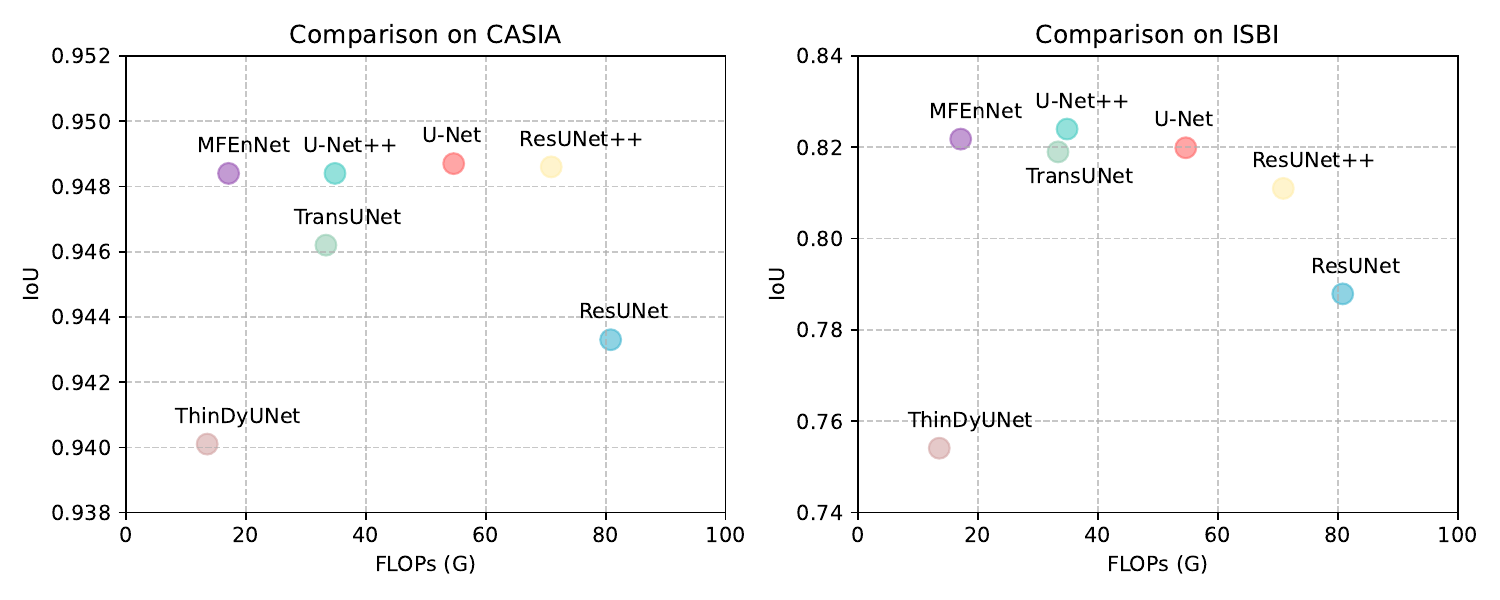}
\caption{The trade-off between accuracy (IoU) vs complexity (FLOPs).}
\label{fig:tradeoff}
\end{figure*}

The emergence of vision transformers (ViTs) \cite{dosovitskiy2021an} has revolutionized image analysis by processing images as patch sequences and using self-attention to capture global context, achieving top performance in computer vision. However, the high computational and memory demands hinder their application in resource-constrained medical imaging settings, such as portable devices and edge systems. To address this, lightweight transformer variants and hybrid architectures have been developed to balance accuracy and efficiency. Notably, MetaFormer \cite{yu2022metaformer} has emerged as a promising architectural abstraction, demonstrating that the core transformer design can be decoupled from self-attention and generalized with alternative token mixers, suggesting that efficient token aggregation mechanisms may serve as viable substitutes for self-attention in medical image segmentation.

This work introduces MFEnNet, a MetaFormer-driven Encoding Network for medical image segmentation that employs pooling token mixers in its encoding backbone. By replacing conventional self-attention with pooling-based operations, MFEnNet achieves global feature aggregation at substantially reduced computational cost. Swish activation is utilized to ensure smoother gradients during backpropagation, mitigating the vanishing gradient issue for more stable training. Spatial pyramid pooling (SPP) \cite{he2015spatial} is strategically implemented at the bottleneck to improve multi-scale feature extraction, effectively capturing both local details and global context across various spatial resolutions. The proposed MFEnNet is evaluated on multiple medical segmentation benchmarks, including CASIA \cite{CASIA-Iris} and ISBI \cite{gutman2016skin}, to assess both segmentation accuracy and computational efficiency. The trade-off between these two aspects is illustrated in Figure \ref{fig:tradeoff}, showing that MFEnNet attains comparable accuracy against state-of-the-art models with a substantially lower computational cost. In a nutshell, the main contributions of this work are summarized as follows: 1) MetaFormer is adapted for medical image segmentation, demonstrating its effectiveness as an alternative to traditional CNN and self-attention–based transformer backbones; 2) self-attention is replaced with pooling operation in transformer blocks, enabling efficient feature aggregation while maintaining competitive segmentation accuracy; 3) the proposed MFEnNet is evaluated on multiple medical benchmarks, achieving state-of-the-art accuracy with significantly lower computational cost, thereby improving suitability for resource-constrained applications.

\section{Related Work}
\label{sec:relatedwork}

Early advances in medical image segmentation were driven by U-Net \cite{ronneberger2015u}, which introduced a symmetric encoder–decoder architecture and became the de-facto baseline for many vision tasks \cite{tran2019robust,tran2022novel}. Building on U-Net, numerous variants, such as ResUNet \cite{zhang2018road}, UNet++ \cite{zhou2018unet++}, and ResUNet++ \cite{jha2019resunet++}, incorporated residual learning, redesigned the skip connections, and added attention mechanisms to further refine feature representation and improve segmentation accuracy.

Motivated by ViTs' ability to capture long-range dependencies, various works integrated CNN backbones with transformer encoders, allowing models to exploit global context while retaining spatial precision. For instance, TransUNet \cite{chen2024transunet} employs a CNN to extract low-level features, which are tokenized and processed by a transformer encoder before being decoded in a U-Net style. Swin-Unet \cite{cao2022swin} replaces convolutional encoders with hierarchical windowed transformers that jointly model local and global representations. TransFuse \cite{zhang2021transfuse} explicitly fuses parallel CNN and transformer streams to leverage complementary cues. Beyond architectural design, several approaches have been introduced to address the unique statistics and limited-data regimes of medical datasets; for example, gated/axial attention \cite{valanarasu2021medical} and local-global parallel aggregation \cite{liu2022phtrans}. While self-attention in ViTs provides strong representational capacity, its computational complexity poses challenges for high-resolution medical images and resource-constrained deployment. This has spurred research into more efficient token-mixing strategies. Notably, MetaFormer \cite{yu2022metaformer} revealed that much of a transformer's effectiveness derives from its general architectural design rather than the attention mechanism itself, showing that the token mixer can be replaced with simpler alternatives without significant loss in performance.

Inspired by MetaFormer, the proposed framework integrates its principles into medical image segmentation, employing pooling transformer blocks to achieve efficient global context aggregation with markedly reduced computational overhead compared to conventional self-attention–based models.

\begin{figure*}[t]
\centering
\includegraphics[width=1.0\textwidth]{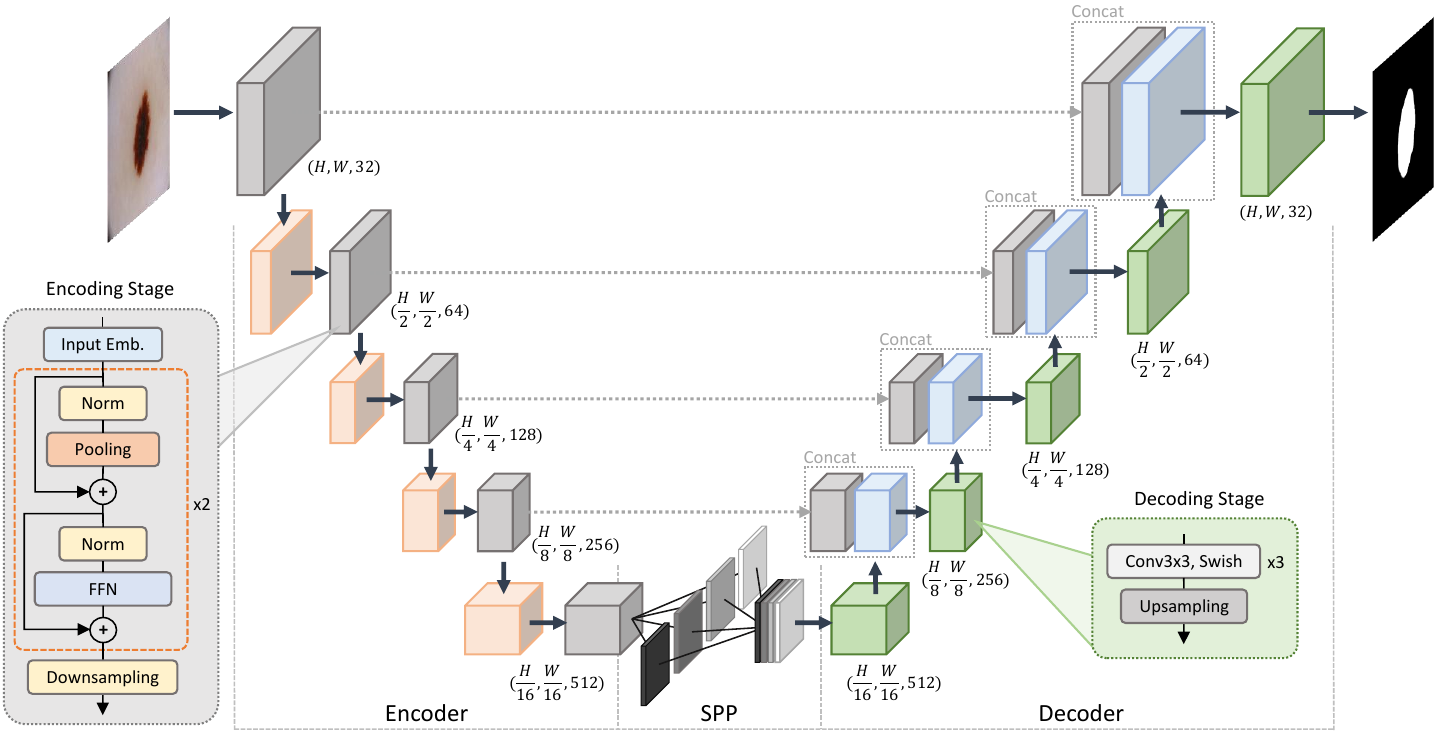}
\caption{The proposed network.}
\label{fig:network}
\end{figure*}

\section{Methodology}
\label{sec:methodology}

The proposed MFEnNet integrates a MetaFormer-inspired block into the U-Net encoder for global context modeling, incorporates an SPP module at the bottleneck for multi-scale feature aggregation, and employs Swish activation to improve gradient flow, collectively achieving strong performance with reduced computational cost.

\subsection{MetaFormer-driven Encoder}
\label{subsec:mfe}

In a standard ViT block, the input feature map $X$ is partitioned into patches and embedded into a token sequence:
\begin{equation}
Z = \mathrm{InputEmb}(X),
\label{eq2}
\end{equation}
where $Z \in \mathbb{R}^{N \times C}$ denotes $N$ tokens with $C$ channels. The sequence $Z$ is then processed through two sub-blocks. The first performs token mixing:
\begin{equation}
Z_1 = Z + \mathrm{TokenMixer}(\mathrm{Norm}(Z)),
\label{eq3}
\end{equation}
and the second applies a feed-forward transformation:
\begin{equation}
Z_2 = Z_1 + \mathrm{FFN}(\mathrm{Norm}(Z_1)).
\label{eq4}
\end{equation}
Here, $\mathrm{TokenMixer}(.)$ enables interaction among tokens (typically via self-attention), $\mathrm{Norm}(.)$ denotes normalization to stabilize training, and $\mathrm{FFN}(.)$ is a two-layer feed-forward network with non-linear activation.

In the proposed model, the encoder is reformulated following the MetaFormer paradigm \cite{yu2022metaformer}. Specifically, the token mixer is replaced with a pooling operator, yielding the modified first sub-block:
\begin{equation}
Z_1 = Z + \mathrm{Pooling}(\mathrm{Norm}(Z)),
\label{eq3d}
\end{equation}
while the second sub-block remains unchanged. The $\mathrm{FFN}(.)$ employs two fully connected layers with an expansion ratio $r=4$ and Swish activation $\sigma$, and each sub-block is equipped with a skip connection to facilitate information flow. For input embedding, a $3 \times 3$ convolution is used. The encoding stage of the proposed framework is illustrated in Figure \ref{fig:network}. This design preserves the long-range dependency modeling characteristic of ViT architectures while substantially reducing memory usage and computational requirements.

\begin{figure*}[t]
\centering
\includegraphics[width=1.0\textwidth]{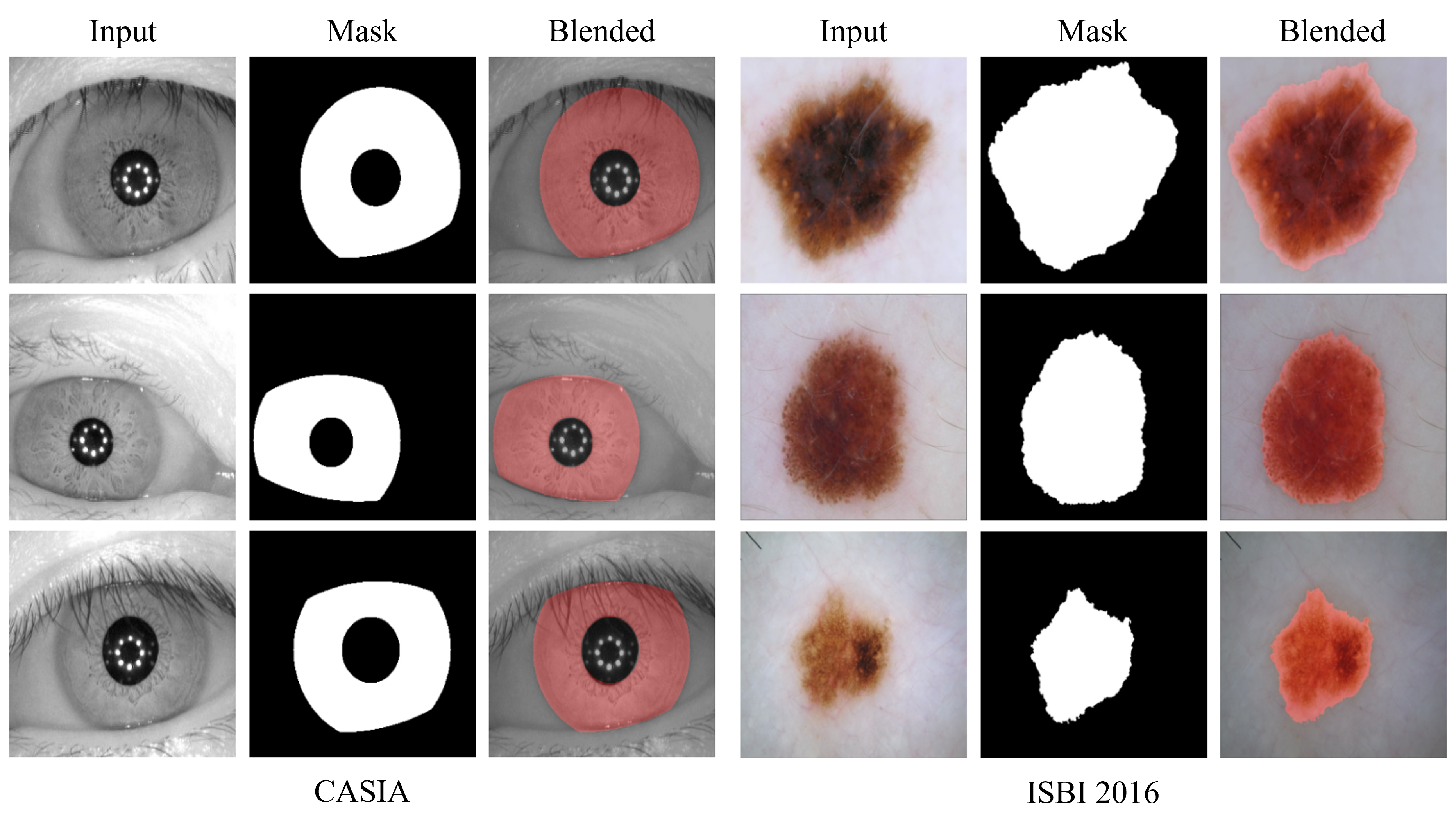}
\caption{Examples of data used in the experiment.}
\label{fig:dataset_examples}
\end{figure*}

\subsection{MFEnNet}
\label{subsec:mfe_unet}

The proposed MFEnNet is constructed on top of a vanilla U-Net architecture, consisting of an encoder–decoder structure with long skip connections to preserve fine-grained spatial information across scales, as shown in Figure \ref{fig:network}. In the encoder, each level begins with an input embedding layer, followed by a stack of encoding blocks where pooling is employed as the token mixing strategy. Down-sampling is performed using max pooling to progressively enlarge the receptive field while compressing spatial resolution. The decoder mirrors the U-Net design, gradually reconstructing high-resolution feature maps through up-sampling and concatenation with encoder features from corresponding levels. Unlike the standard U-Net, however, we replace ReLU with the Swish activation function, which provides smoother gradients and has been shown to improve optimization stability and representational power. Drawing inspiration from notable image-to-image translation works \cite{tran2025encoder,zhao2017pyramid}, we incorporate an SPP module at the bottleneck. This module aggregates multi-scale contextual information by applying pooling operations over regions of varying sizes, enabling the network to capture both global semantics and fine local structures. The network is structured across five stages (including the input/output stage and four down/up-sampling steps) with feature map dimensions of $H \times W \times 32$, $\frac{H}{2} \times \frac{W}{2} \times 64$, $\frac{H}{4} \times \frac{W}{4} \times 128$, $\frac{H}{8} \times \frac{W}{8} \times 256$, and $\frac{H}{16} \times \frac{W}{16} \times 512$, where $H$ and $W$ represent the input image's height and width, respectively. MetaFormer blocks are employed solely in the encoder, as ViT-based blocks are particularly effective at modeling long-range dependencies in the input sequence, which is critical for contextual understanding in the encoding phase \cite{chen2024transunet,tran2025distilled}. At the output layer, a plain $1 \times 1$ convolutional layer generates the final segmentation map.


\section{Experiments}
\label{sec:experiments}

\setlength{\tabcolsep}{8pt} 
\begin{table}[t]
\centering
\caption{Comparisons of various methods in terms of accuracy and complexity.}
\resizebox{1.0\textwidth}{!}{
\begin{tabular}{l|cc|cc|cc}
\toprule
\multirow{2}{*}{\text{Model}} & \multicolumn{2}{c|}{CASIA} & \multicolumn{2}{c|}{ISBI} & \multicolumn{2}{c}{Complexity} \\
\cmidrule{2-7}
 & IoU & Dice & IoU & Dice & Params (M) & FLOPs (G) \\
\midrule
U-Net \cite{ronneberger2015u} & 0.9487 & 0.9734 & 0.8199 & 0.8911 & 31.04 & 54.66 \\
ResUNet \cite{zhang2018road} & 0.9433 & 0.9705 & 0.7879 & 0.8675 & 13.04 & 80.83 \\
U-Net++ \cite{zhou2018unet++} & 0.9484 & 0.9730 & 0.8238 & 0.8947 & 9.16 & 34.87 \\
TransUNet \cite{chen2024transunet} & 0.9462 & 0.9720 & 0.8195 & 0.8894 & 3.63 & 33.36 \\
ResUNet++ \cite{jha2019resunet++} & 0.9486 & 0.9733 & 0.8110 & 0.8854 & 14.48 & 70.92 \\
ThinDyUNet \cite{kim2025semantic} & 0.9401 & 0.9688 & 0.7541 & 0.8428 & 0.81 & 13.56 \\
\midrule
MFEnNet (ours) \quad & 0.9484 & 0.9732 & 0.8218 & 0.8913 & 11.14 & 17.13 \\
\bottomrule
\end{tabular}
}
\label{tab:performance_comparison}
\end{table}

\subsection{Experimental Settings}

\textbf{Datasets:} The experiments utilized two publicly available datasets: CASIA Iris Interval (CASIA) \cite{CASIA-Iris} and ISBI 2016 (ISBI) \cite{gutman2016skin}. The CASIA dataset comprises 2,639 iris images obtained from 395 eyes of 249 subjects using a consistent sensor. This dataset was partitioned randomly into 80\% for training and 20\% for validation. On the other hand, the ISBI dataset, designed for skin lesion segmentation, comprises 900 training images and 379 test images for evaluation. Representative samples from both datasets are illustrated in Figure~\ref{fig:dataset_examples}.

\textbf{Experimental Setup:} All experiments were conducted on a Linux-based system equipped with NVIDIA Tesla T4 GPUs. The proposed framework was implemented using the PyTorch library and trained for 50 epochs using the Adam optimizer with a batch size of 16 and a learning rate of $10^{-4}$. Binary cross-entropy (BCE) was employed as the loss function. Input images were resized to $256 \times 256$ pixels. To enhance model robustness, data augmentation techniques, including random flipping and random cropping, were applied during training. Quantitative performance was evaluated using the Intersection over Union (IoU) and Dice Coefficient, while computational efficiency was assessed via the number of trainable parameters (Params, M) and floating-point operations (FLOPs, G) for a $256 \times 256$ input. 

\textbf{Baselines:} The proposed network has been compared against state-of-the-art semantic segmentation models, representing a broad spectrum of approaches with both CNN-based and transformer-based architectures, ensuring a comprehensive and fair comparison. These models include U-Net~\cite{ronneberger2015u}, U-Net++~\cite{zhou2018unet++}, ResUNet~\cite{zhang2018road}, ResUNet++~\cite{jha2019resunet++}, TransUNet~\cite{chen2024transunet}, and ThinDyUNet~\cite{kim2025semantic}. These baselines enable a comprehensive comparison of performance and efficiency.

\begin{figure}[t]
\centering
\includegraphics[width=1.0\textwidth]{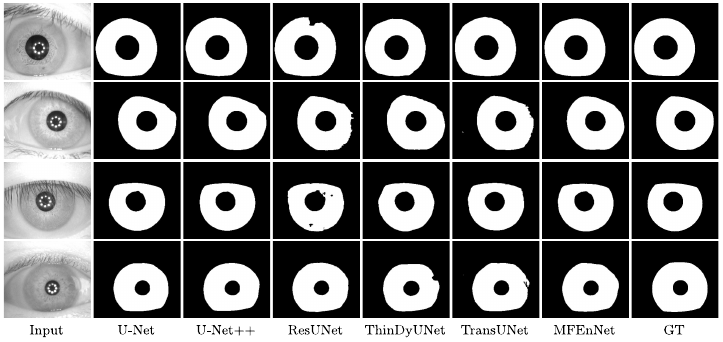}
\caption{Typical visual comparisons of various approaches on CASIA data.}
\label{fig:casia}
\end{figure}

\subsection{Quantitative Analysis}

Table \ref{tab:performance_comparison} reports a comparative evaluation of the proposed MFEnNet against representative CNN- and transformer-based models on the CASIA and ISBI benchmarks, with respect to segmentation accuracy and model complexity.

Across both datasets, U-Net and its variants, such as ResUNet, U-Net++, and ResUNet++, consistently demonstrate strong segmentation performance, reaffirming the robustness of convolutional encoder–decoder architectures. Among these, U-Net++ achieves the highest performance on the ISBI dataset, with an IoU of 0.8238 and a Dice coefficient of 0.8947, highlighting the effectiveness of dense skip connections in enhancing feature fusion. However, this improved accuracy comes at the cost of increased computational complexity, particularly when compared to more lightweight models such as ThinDyUNet and MFEnNet. TransUNet, a transformer-based approach, also yields competitive results with a small model size, demonstrating the advantages of global context modeling. Despite this, such models typically involve significantly higher computational demands; for example, TransUNet has a relatively modest parameter count (3.63M) but requires 33.36 GFLOPs. Similarly, ResUNet++ achieves favorable performance but with 14.46M parameters and 70.92 GFLOPs, potentially limiting its applicability in resource-constrained environments. In contrast, the proposed MFEnNet achieves a favorable balance between accuracy and efficiency. On CASIA, it attains an IoU of 0.9484 and a Dice score of 0.9732, closely matching the best-performing baselines. On ISBI, it achieves an IoU of 0.8218 and a Dice score of 0.8913, performing on par with U-Net++ while substantially reducing FLOPs. Specifically, MFEnNet requires only 11.14M parameters and 17.13G FLOPs, representing a 64\% reduction in parameters and nearly 68\% reduction in FLOPs relative to U-Net, while maintaining comparable segmentation accuracy.

These findings highlight two important insights. First, transformer-based token mixing, when implemented efficiently through MetaFormer-inspired pooling blocks, can retain the representational advantages of global context modeling without incurring prohibitive computational costs. Second, the integration of multi-scale aggregation (via SPP) and smooth optimization dynamics (via Swish activation) further contributes to stable training and competitive segmentation outcomes. Overall, the proposed MFEnNet delivers state-of-the-art segmentation accuracy with substantially lower complexity, thus enhancing its suitability for resource-constrained medical applications.

\begin{figure}[t]
\centering
\includegraphics[width=1.0\textwidth]{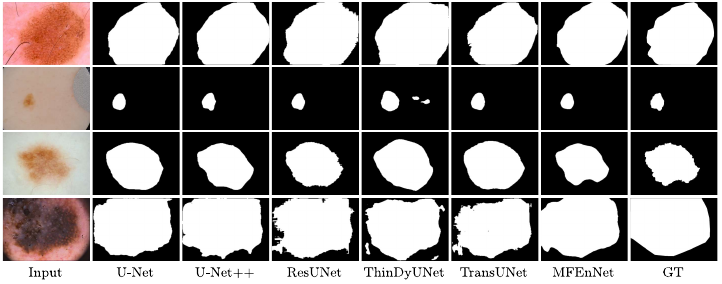}
\caption{Typical visual comparisons of various approaches on ISBI data.}
\label{fig:isbi}
\end{figure}

\subsection{Qualitative Analysis}

Figure \ref{fig:casia} and Figure \ref{fig:isbi} present visual comparisons of segmentation results produced by state-of-the-art models, including U-Net, U-Net++, ThinDyUNet, ResUNet, TransUNet, and the proposed MFEnNet.

As can be observed from Figure \ref{fig:casia}, which illustrates segmentation outcomes for iris images from the CASIA dataset, CNN-based methods, such as U-Net and U-Net++, produce reasonable results with sharp boundaries. ResUNet often yields slight boundary distortions and suffers from small fragmented regions and noisy predictions, while ThinDyUNet under-segments in multiple cases due to its reduced representational capacity. TransUNet, leveraging self-attention, improves boundary smoothness but still introduces error and local artifacts under varying illumination conditions. In comparison, MFEnNet yields masks that are visually closest to the ground truth, with clean and continuous contours around both iris and pupil regions. 

On the other hand, Figure \ref{fig:isbi} presents results on skin lesion images from the ISBI dataset, which pose additional challenges due to irregular lesion shapes, blurred boundaries, and variable lesion sizes. U-Net and U-Net++ tend to over-segment, leading to masks that extend beyond the true lesion region. ResUNet introduces structural inconsistencies, while ThinDyUNet misses finer lesion details, producing incomplete masks. TransUNet performs better in capturing lesion extent but generates uneven contours in complex cases, especially when background textures resemble lesion patterns. In contrast, MFEnNet consistently delineates lesion boundaries with high fidelity, capturing irregular edges while avoiding over-segmentation. Notably, in small-lesion cases, the proposed MFEnNet maintains precise localization, whereas competing models either under-segment or introduce false positives.

\section{Conclusions}
\label{sec:conclusions}

In this work, we present MFEnNet, an efficient medical image segmentation framework that leverages a MetaFormer-inspired encoder to balance segmentation accuracy and computational efficiency. By replacing conventional self-attention with pooling transformer blocks, MFEnNet effectively aggregates global contextual information while maintaining low complexity. The integration of a spatial pyramid pooling (SPP) module at the bottleneck further enhances multi-scale feature representation, and the use of Swish activation facilitates stable optimization and improved gradient flow. Evaluations on benchmark datasets, including CASIA and ISBI, demonstrate that MFEnNet achieves competitive segmentation accuracy against state-of-the-art methods, while substantially reducing computational cost. Qualitative comparisons further confirm that MFEnNet delivers segmentation masks with cleaner boundaries and more faithful structural representation, highlighting its robustness across diverse imaging scenarios such as iris boundary extraction and skin lesion delineation. By offering a strong trade-off between performance and efficiency, MFEnNet shows promise for deployment in resource-constrained medical applications. 

%
%

\bibliographystyle{unsrt}
\bibliography{ref}

\end{document}